\documentclass[preprint,12pt]{revtex4-1}
\usepackage{amsmath}
\usepackage{amssymb}
\usepackage{mathtools}
\usepackage{graphicx}
\usepackage{geometry}
\usepackage{caption}
\usepackage{float}
\usepackage{subcaption}
\usepackage{hyperref}
\usepackage{tocloft}
\usepackage{verbatim}
\usepackage{comment}
\usepackage{titlesec}
\usepackage{slashed}
\usepackage{bm}
\hypersetup{colorlinks=true,linkcolor=blue}

\newcommand{\Hom}{\text{\Hom}}

\DeclareMathSymbol{:}{\mathord}{operators}{"3A}

\DeclarePairedDelimiterX\braket[2]{\langle}{\rangle}{#1 \delimsize\vert #2}
\DeclarePairedDelimiterX\ketbra[2]{\delimsize\vert}{\delimsize\vert}{#1
\rangle\langle
 #2}

\begin{document}
 
\title{Regularized Stokeslets lines suitable for slender bodies in viscous flow}
\date{\today}
\author{Boan Zhao}
\affiliation{ Department of Applied Mathematics and Theoretical Physics,University of Cambridge, Wilberforce Road,Cambridge CB3 0WA, United Kingdom;}
\author{Lyndon Koens\footnote{lyndon.koens@mq.edu.au}}
\affiliation{ Department of Mathematics and Statistics, Macquarie University, 192 Balaclava Rd, Macquarie Park, NSW 2113, Australia}
\begin{abstract}
Slender-body approximations have been successfully used to explain many phenomena in low-Reynolds number fluid mechanics. These approximations typically use a line of singularity solutions to represent the flow. These singularities can be difficult to implement numerically because they diverge at their origin. Hence people have regularized these singularities to overcome this issue. This regularization blurs the force over a small blob therefore removing the divergent behaviour. However it is unclear how best to regularize the singularities to minimize errors. In this paper we investigate if a line of regularized Stokeslets can describe the flow around a slender body. { This is achieved by comparing the asymptotic behaviour of the flow from the line of regularized Stokeslets with the results from slender-body theory.} We find that the flow far from the body can be captured if the regularization parameter is proportional to the radius of the slender body. { This is consistent with what is assumed in numerical simulations and provides a choice for the proportionality constant.} However more stringent requirements must be placed on the regularization blob to capture the near field flow outside a slender body. This inability to replicate the local behaviour indicates that many regularizations cannot satisfy the non-slip boundary conditions on the bodies surface { to leading order}, with one of the most commonly used regularizations showing an angular dependency of velocity along any cross section. This problem can be overcome with compactly supported blobs { and we construct one such example blob which could be effectively used to simulate the flow around a slender body}.
\end{abstract}
\maketitle
\def\v{\vspace{2cm}}

\section{Introduction}
Slender bodies immersed in fluids are frequently
studied in biology, polymer mechanics and colloids. These bodies typically have arc length much larger than their radius and can be difficult to resolve numerically. Let $a$ denote the ratio of the radius of a slender body to its length. In the small $a$ limit, asymptotic theories have been developed for the  flow around slender bodies \cite{Cox, 1976, Batchelor2006, Keller1976a, Johnson1979, Gotz2000, Koens2018, Andersson2020, Barta1988, Koens2021a}. These theories are called slender-body theories (SBTs), and expand the system in powers of $a$ or $1/\ln a$. Algebraically accurate SBTs have proven to be very accurate and give exact results for prolate spheroids \cite{Batchelor2006, Johnson1979, Koens2018, Kim2004a, Das2018, Higdon2006, BARTA2006, Cummins2018, Rodenborn2013a} while the logarithmically accurate SBTs are useful for analytical estimations \cite{GRAY1955, Chattopadhyay2006, Koens2018a, Zhang2019c, BECKER2003, Tatulea-Codrean2021, Waszkiewicz2021a, Man2017a}. Many of these theories place the singularity solutions of Stokes equations along the centreline of the body. This singularity representation diverges on the centreline of the body and so can become difficult to implement numerically.

One approach to overcome these problems is to regularize the singularities by distributing the force over a small blob \cite{Cortez2001, Cortez2005, Zhao2019a}. This regularization of singularities was introduced by Cortez \cite{Cortez2001} with applications to the boundary integral representation for the flow \cite{Cortez2005, Smith2009, Gallagher2021, Nguyen2014}.  It has since been extended to describe the flow around slender bodies through different regularized slender-body theories \cite{Walker, Cortez2012, Cortez2018, Buchmann2018, Martindale2016, HoaNguyen2014}.

{ Two major versions of these }regularized SBTs exist: { one in which both regularised point forces and source dipoles are distributed over the centreline \cite{Cortez2012, Walker, HoaNguyen2014} and one which} consists of placing a series of regularized point forces along the centreline of the body \cite{Nguyen2019, Buchmann2018, Martindale2016, Bouzarth2020, Olson2011, Montenegro-Johnson2016}. { The former uses asymptotic methods to match the boundary conditions like in SBT while the latter assumes} that the regularized singularities will produce the correct flow outside the slender body with a suitable choice of regularization parameter. This is aruged from unit analysis but has not been mathematically proven. { Here we investigate the accuracy of the latter representation but note that a recent analysis has looked into the accuracy of the former \cite{Ohm2021}.} 

In this paper we compare the { asymptotic } flows around a line of regularized singularities to the flow around a slender body { as predicted by slender-body theory.} We show that the asymptotic flow far from the body can be captured by any regularization provided the regularization parameter satisfies a specific equation that depends on the shape of the blob. { This justifies the linear relationship between $\epsilon$ and the thickness of the slender body which is typically assumed in numerical simulations.} We also show that the flow close to the slender body can only be { asymptotically} captured for specific types of blobs. As such not all regularization choices are capable of satisfying the non-slip boundary conditions on the body and so cannot capture the behaviour of the slender body itself. This is an issue for a commonly used power-law decaying blob but not for compactly supported blobs. { We construct one such compact blob which could be effectively used for numerical simulations.}

In section \ref{sec:background} we briefly review Stokes flow with a focus on classical and regularized singularity solutions and slender-body theories. Section \ref{sec:force on the body} then identifies the conditions under which the { asymptotic} flow far from a line of regularized Stokeslets is the same as that far from a slender body before we identify the conditions necessary to capture the near flow in Sec.~ \ref{sec:velocity near the body}. Finally we look at how these conditions apply to typical blob types in Sec.~\ref{sec:examples} before we conclude the paper.

\section{Background into viscous flows around slender bodies} \label{sec:background}

\subsection{Stokes flow and classical singularity solutions}
At microscopic scales the ratio of fluid inertia to fluid viscosity, the Reynolds number, tends to be very small. In these cases the behaviour of the resultant slow viscous flow can be accurately described by the incompressible Stokes equations \cite{Kim2005,  LAMB1932, Pozrikidis1992}
\begin{eqnarray}
-\nabla p+\mu \nabla^2\boldsymbol{u} +\boldsymbol{f} &=& 0
,\\
\nabla\cdot\boldsymbol{u} &=& 0,
\end{eqnarray}
where $p$ is the pressure of the fluid, $\boldsymbol{u}$ is the velocity of the fluid and $\boldsymbol{f}$ is the force per unit volume on the fluid. Throughout we will set $\mu=1$ without any loss of generality. Though these equations are linear and time independent, they display a strong dependence on the geometry of the system considered. As a result analytical solutions to these equations are rare except in very specific geometries \cite{Kim2005, LAMB1932, Pozrikidis1992}. Hence, for a typical problem, they must be evaluated numerically or asymptotically. Both of these approaches often use the point force solution to the flow.

The fluid flow from a point force, $\boldsymbol{f = F\delta{(r)}}$ { where $\boldsymbol{\delta{(r)}}$ is the three dimensional Dirac delta function supported at the origin}, in a viscous fluid is called the Stokeslet and, in free space, generates the flow
\begin{eqnarray}
\boldsymbol{u}(\boldsymbol{r}) = \boldsymbol{S}(\boldsymbol{r})\cdot\boldsymbol{F}
&=& \left(\frac{r^2\boldsymbol{I}+\boldsymbol{rr}}{8\pi 
r^3}\right)\cdot\boldsymbol{F},\\
p(\boldsymbol{r}) =
\boldsymbol{P}(\boldsymbol{r})\cdot\boldsymbol{F} &=&
\left(\frac{\boldsymbol{r}}{4\pi r^3}\right)\cdot\boldsymbol{F},
\end{eqnarray}
where $\boldsymbol{r}$ is the position vector, $r= |\boldsymbol{r}|$ is its norm,  $\boldsymbol{S}(\boldsymbol{r})$ is the Oseen tensor, $\boldsymbol{I}$ is the identity tensor and all tensor product
symbols are omitted. The Stokeslet is the Green's function of the flow and satisfies the Stokes equations everywhere outside the origin at which point it diverges. As a result it can be used to solved the Stokes equations in two different ways: the boundary integral representation and the representation by fundamental singularities \cite{Kim2005, Pozrikidis1992}. In the boundary integral representation, the Green's function nature of the Stokeslet is used to convert the equations into an integral over the boundary of the domain. These integrals can then be discretized to solve for the unknowns in the problem in a method called the boundary element method (for more information on this method see Ref.~\cite{Pozrikidis1992}).

The representation by fundamental singularity method, on the other hand, seeks to place the Stokeslet and its derivatives outside of the flow region (ie. within the body) such that the boundary conditions are satisfied \cite{Chwang2006}. If such a distribution can be found, it a guaranteed to be the solution due to the uniqueness of the Stokes equations. In general the specific singularities needed depends on the geometry of the problem, the region they are distributed over and the motion of the body. However, due to the bi-harmonic nature of the Stokes equations, singularity representations involving the Stokeslet are consistently found to occur in combination with another singularity solution called the source dipole. The source dipole is proportional to the Laplacian of the Stokeslets, and generates the flow
\begin{eqnarray}
\boldsymbol{u}(\boldsymbol{r}) = \boldsymbol{D}(\boldsymbol{r})\cdot\boldsymbol{A}
&=& \frac{r^2\boldsymbol{I}-3\boldsymbol{rr}}{8\pi
r^5}\cdot\boldsymbol{A},\\
p(\boldsymbol{r}) &=&0,
\end{eqnarray}
where  $\boldsymbol{A}$ is the strength of the source dipole, and $\boldsymbol{D(r)} = \nabla^2\boldsymbol{S(r)}/2$ is the source dipole tensor. Similarly to the Stokeslet this fundamental singularity satisfies the Stokes equations, everywhere except at the origin where it diverges.

\subsection{Regularized singularity solutions} 

The divergent nature of the Stokeslets requires special care when integrating over or determining the flow close to one. However this integration is necessary for the boundary integral representation and the nearby flow is often required when constructing representations using fundamental singularities. Cortez proposed the regularization of these singularities to overcome this issue \cite{Cortez2001}. This was achieved by replacing the Dirac delta function in the point force representation by a suitable mollifier (hereafter called a blob), $f_\epsilon(\boldsymbol{r}) = \epsilon^{-3}f(\boldsymbol{r}/\epsilon)$ where $ \iiint f_{\epsilon}(\boldsymbol{r}) \,d\boldsymbol{r} = 1$, that depends on some regularization parameter $\epsilon$. As $\epsilon \to 0$ these blobs become the Dirac delta function and so the flow from a Stokeslet is recovered. The flow from a specific blob can be found by solving the forced Stokes equations or represented as a convolution with the relevant singularity. For example the regularized Oseen tensor can be generally expressed as

\begin{equation}
\boldsymbol{S}^\epsilon(\boldsymbol{r}) = \boldsymbol{S}*f_\epsilon = \int \boldsymbol{S(r -
r')}f_\epsilon(\boldsymbol{r'})d\boldsymbol{r}'
\end{equation}
but a popular regularization takes the form 

\begin{equation}
\boldsymbol{S}^\epsilon(\boldsymbol{r}) = \frac{\boldsymbol{I}(r^2 + 2\epsilon^2) +
\boldsymbol{r}\boldsymbol{r}}{8\pi r_\epsilon^3}, \qquad r_\epsilon = \sqrt{r^2 +
\epsilon^2} \label{pop}
\end{equation}
when the blob is $f_{\epsilon}(r) = 15 \epsilon^4/(8\pi r_\epsilon^7)$. The flow from these tensors define the regularized Stokeslet.

The substitution of this regularized Stokeslet into the boundary integral representation develops the regularized boundary element method. This method overcomes any singularity issues caused by the Stokeslet but introduces an error related to the regularization parameter $\epsilon$ \cite{Cortez2005}. Careful analysis of this error showed that for a spherically symmetric blobs the additional error to the boundary integral representation proportional to $\epsilon^2$ \cite{Nguyen2014}.

This error can be understood through the additional flows generated by the regularization process \cite{Zhao2019a}. We previously showed that when $r  \gg \epsilon$ the regulalarized Stokeslet $S^\epsilon$ can be written as an infinite sum of singularity solutions and when $r \ll \epsilon$, $\boldsymbol{S^\epsilon}$ becomes isotropic. Furthermore when $f$ is spherically symmetric, the regularized Oseen tensor can be { expressed exactly \cite{Zhao2019a} as}

\begin{equation}\label{Stokeslet_r}
\boldsymbol{S}^\epsilon(\boldsymbol{r}) =
\boldsymbol{S}(\boldsymbol{r})\int_0^{r/\epsilon}4\pi t^2f(t)dt +
\epsilon^2\boldsymbol{D}(\boldsymbol{r})\int_0^{r/\epsilon}\frac{4\pi}{3}
t^4f(t)dt
+\frac{2\boldsymbol{I}}{3\epsilon}\int_{r/\epsilon}^\infty t
f(t)dt,
\end{equation}
where we have now expressed it in terms of fundamental singularity solutions to the Stokes equations. The above representation shows that the flow far from a regularized Stokeslet can always be expressed as that of a Stokeslet and a source dipole proportional to $\epsilon^2$ regardless of the choice of blob. Hence a spherically symmetric regularizations generally produce additional flows at order $\epsilon^2$.

This $\epsilon^2$ error can be controlled by keeping $\epsilon$ much smaller than all other lengths within a given problem, thereby allowing the benefits of regularization to far out way any draw backs. As such regularized boundary element simulations have become an increasingly popular tool to explore the behaviour of micro-swimmers \cite{Buchmann2018, Nguyen2019, Bouzarth2020, Olson2011, Montenegro-Johnson2012, Montenegro-Johnson2018, Smith2009}, and have been extended beyond free space singularities \cite{Montenegro-Johnson2012, Ainley2008, Cortez2015}. Such is the appeal of this representation that it has been adapted to the asymptotic representation by fundamental singularities used to determine the flow around a slender body \cite{Walker, Cortez2012, Cortez2018, Martindale2016, HoaNguyen2014, Montenegro-Johnson2016}.

\subsection{Classical slender-body theory} 

The slow viscous hydrodynamics of long thin bodies is important in many systems \cite{1976}. However, this slender-body hydrodynamics can be hard to simulate thanks to the very large aspect ratio $1/a$ (major/minor lengths) of the body. This large aspect ratio causes boundary element methods to typically require high surface resolutions. Hence asymptotic methods, called slender-body theories (SBTs), have been developed to overcome these issue \cite{Cox, 1976, Batchelor2006, Keller1976a, Johnson1979, Gotz2000, Koens2018, Koens2021a}. These theories seek to expand the governing equations in terms of the large aspect ratio $1/a$ to produce simplified models for the hydrodynamics. Broadly SBTs have been divided into two types: those that expand the system in powers of $1/\ln(a)$ \cite{Cox, Batchelor2006} and those that expand the problem in powers of $a$ \cite{ Keller1976a, Johnson1979, Gotz2000, Koens2018}. The former of these is often call resistive force theories and provide useful analytical estimates \cite{GRAY1955, Chattopadhyay2006, Koens2018a, Zhang2019c, BECKER2003, Tatulea-Codrean2021, Waszkiewicz2021a, Man2017a} while the later have been shown to give highly accurate results in numerous situations \cite{Kim2004a, Das2018, Higdon2006, BARTA2006, Cummins2018, Rodenborn2013a}.

\begin{figure}
\includegraphics[height=200pt]{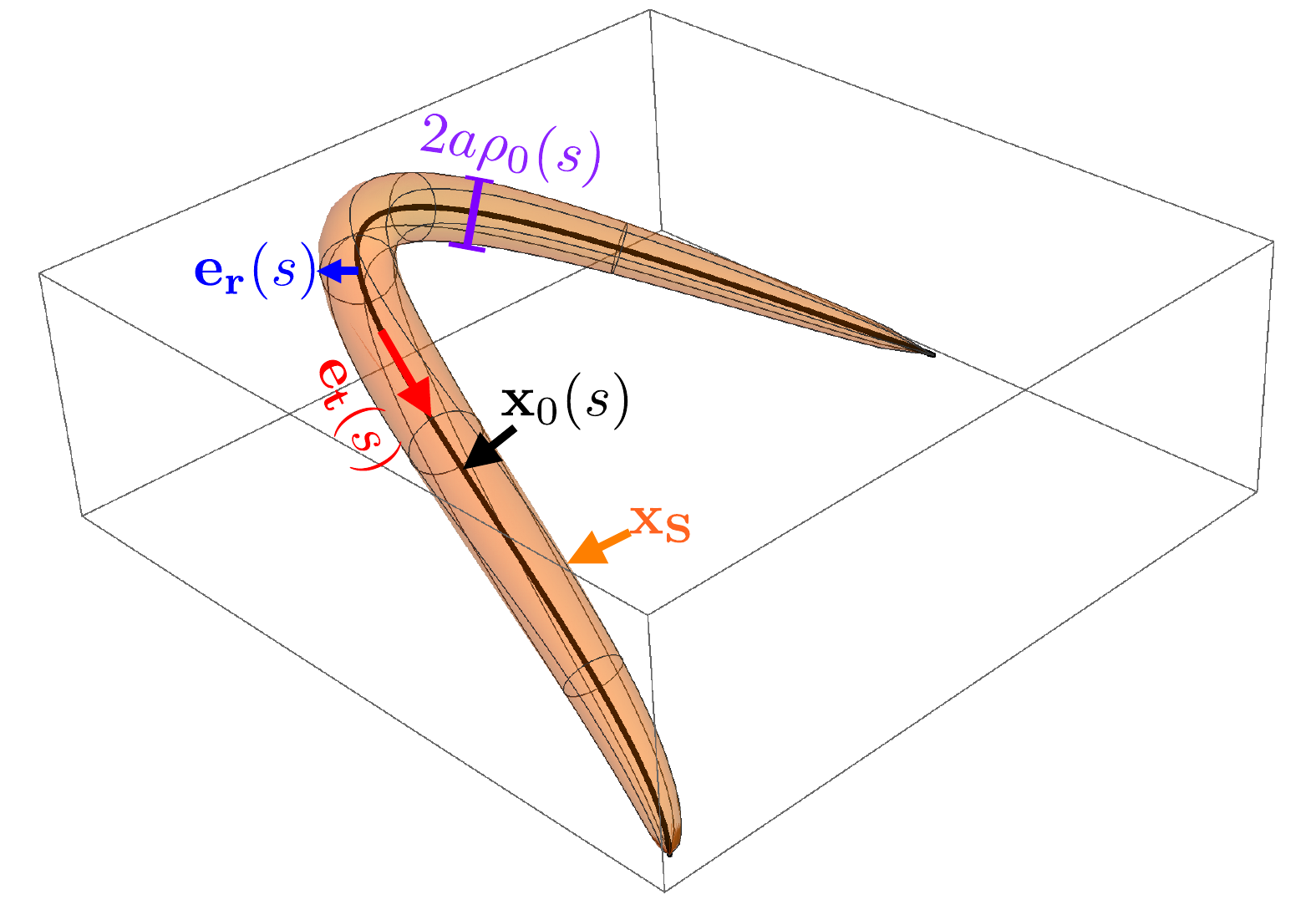}
\caption{Example geometry of a slender body. $\boldsymbol{x}_0(s)$ is the centreline, $\boldsymbol{x_S}$ is the surface, $\boldsymbol{e_t}(s)$ is the tangent to the centreline, $\boldsymbol{e_r}(s)$ is a radial vector perpendicular to the tangent and $2 a \rho_0(s)$ represents the thickness at $s$. } 
\label{figure_slender_body}
\end{figure} 

Probably the most commonly used algebraically accurate SBT was first developed by Keller and Rubinow \cite{Keller1976a}. This version was developed by matching the flow around an infinite cylinder to that of a line of Stokeslets. These equations were later re-derived and extended by Johnson \cite{Johnson1979} and Gotz \cite{Gotz2000}, using the representation by fundamental singularities and matched asymptotic expansions. 
They achieved by this by parametrising the surface of the slender body, $\boldsymbol{x_{S}}(s,\theta)$, as
\begin{equation}
   \boldsymbol{x_{S}}(s,\theta)= \boldsymbol{x_0}(s) + a \boldsymbol{e_r}(s,\theta) \rho_0(s),
\end{equation}
where { $s \in [0,1]$ is the arclength of the centreline, $\boldsymbol{x_0}(s)$ is the centreline of the body}, $0 \leq a \rho_0(s) \leq a$ is the body radius at $s$, $\boldsymbol{e_t} = \dot{\boldsymbol{x_0}}/|\dot{\boldsymbol{x_0}}|$ is the unit tangent vector along the centreline and $\boldsymbol{e_r}(s,\theta) $ is the local radial vector perpendicular to $\boldsymbol{e_t}$ (Fig.~\ref{figure_slender_body}). { The radial vector, $\boldsymbol{e_r}$, is often represented through the Frenet-Serret coordinates as $\boldsymbol{e_r}(s,\theta)  = \cos(\theta) \mathbf{\hat{n}} +\sin(\theta) \mathbf{\hat{b}}$, where $\mathbf{\hat{n}}$ and $\mathbf{\hat{b}}$ are the normal and bi-normal vectors to the centreline respectively.}
Johnson and Gotz assumed that to leading order the flow around such a body could be captured by a system of Stokeslets with strength $\boldsymbol{\alpha}^S(s) $ and source dipoles with strength $\boldsymbol{\beta}^S = \frac{1}{2}a^2\rho_0^2\boldsymbol{\alpha}^S$ placed along the body's centreline, $\boldsymbol{x_0}(s)$. This choice of singularity representation comes from the exact solutions known for a prolate spheroid \cite{Chwang2006} and generates the flow 

\begin{equation}\label{flow_s}
\boldsymbol{u}(\boldsymbol{x}) = \int_0^1 \boldsymbol{S(x-x_0}(t))\cdot\boldsymbol{\alpha}^S(t) +
\boldsymbol{D(x-x_0}(t))\cdot\boldsymbol{\beta}^S(t) dt.
\end{equation}
The uniqueness of Stokes flow tells us that this representation must be the solution if the fluid velocity $\boldsymbol{u}(\boldsymbol{x})$ matches the surface velocity of the body when $\boldsymbol{x} = \boldsymbol{x_{S}}$. They then assumed that the surface velocity, $\boldsymbol{U}(s)$, is uniform along each cross-section (ie. it only depends on the position along the centreline $\boldsymbol{x_0}$) and performed a matched asymptotic expansion for $\boldsymbol{u}(\boldsymbol{x_{S}})$ in the limit $a\to0$ while keeping $s$, and $\theta$ fixed.

This matched asymptotic expansion divided the system into an outer and and inner region{, and expands the behaviour in each region in the limit $a\to 0$. The different regions are then matched together using the Van Dyke's matching rule and a composite representation formed \cite{Hinch1991}. We will provide a summary of the major results of this expansion here and refer readers to the thesis of Gotz \cite{Gotz2000} for the full derivation. In the inner region $\xi=(t-s)/a$ is held constant and the integral kernels expand to}  
\begin{eqnarray}
 \boldsymbol{S}^{(i)}(\boldsymbol{x_{S} -
x_0}(t)) &=& \frac{\boldsymbol{I}}{8\pi a\delta_0} +
 \frac{\rho_0^2\boldsymbol{e_re_r}-\rho_0\xi(\boldsymbol{e_re_t+e_te_r}) + \xi^2\boldsymbol{e_te_t}}{8\pi a\delta_0^3} + O(1),  \label{inner_lim_s1}\\
\boldsymbol{D}^{(i)}(\boldsymbol{x_{S} - x_0}(t)) &=& \frac{\boldsymbol{I}}{8\pi a^3\delta_0^3} - 3 \frac{\rho_0^2\boldsymbol{e_re_r}-\rho_0\xi(\boldsymbol{e_re_t+e_te_r}) + \xi^2\boldsymbol{e_te_t}}{8\pi a^3\delta_0^5} \label{inner_lim_s2} + O(a^{-2}), \notag \\ 
\\
\boldsymbol{\alpha}^S(t) &=&
\boldsymbol{\alpha}^S(s) +O(a),
\end{eqnarray}
where { the superscript $(i)$ denotes the inner region expanded form and} $\delta_0(t,s,a) = \sqrt{\xi^2 + \rho_0^2(s)}$. Similarly, the outer limit fixes $t$ and { produces the expanded forms}
\begin{eqnarray}
\boldsymbol{S}^{(o)}(\boldsymbol{x_{S} -
x_0}(t))\cdot\boldsymbol{\alpha}^S(t) &=&
\frac{1}{8\pi}\frac{\boldsymbol{\alpha}^S(t)}{R_0}+\frac{1}{8\pi}
\frac{\boldsymbol{R_0R_0}\boldsymbol{\alpha}^S(t)}{R_0^3} +O(a), \label{outer1}\\
\boldsymbol{D}^{(o)}(\boldsymbol{x_{S} -
x_0}(t))\cdot\boldsymbol{\beta}^S(t) &=& \mathbf{0} +O(a), \label{outer2}
\end{eqnarray}
where  { the superscript $(o)$ denotes the outer region expanded term,} $\boldsymbol{R_0}(t,s) = \boldsymbol{x_0}(s) - \boldsymbol{x_0}(t)$ and its norm is $R_0 = |\boldsymbol{R}_0|$. The common part of these two asymptotic forms is found by expanding as the outer from in terms of the inner regions variables, or equivalently the inner form in terms of the outer regions variables \cite{Hinch1991} and is given by
{
\begin{eqnarray}
\boldsymbol{S^{(o)\in(i)}(x_{S}-x_0}(t))\cdot\boldsymbol{\alpha}^S(t) = \boldsymbol{S^{(i)\in(o)}(x_{S}-x_0}(t))\cdot\boldsymbol{\alpha}^S(t) &=& \frac{\boldsymbol{\alpha}^S(s)}{8\pi|s-t|} + 
\frac{\boldsymbol{e_te_t}\cdot\boldsymbol{\alpha}^S(s)}{8\pi|s-t|}+O(a), \\
\boldsymbol{D^{(o)\in(i)}(x_{S}-x_0}(t))\cdot\boldsymbol{\beta}^S(t) = \boldsymbol{D^{(i)\in(o)}(x_{S}-x_0}(t))\cdot\boldsymbol{\beta}^S(t) &=& \mathbf{0} +O(a),
\end{eqnarray}}
{ 
where the superscript $(o)\in(i)$ denote the outer asymptotic representations, Eqs.~\eqref{outer1} and \eqref{outer2}, expanded in the inner region variables $\xi$, and  $(i)\in(o)$ denote the inner asymptotic representations, Eqs.~\eqref{inner_lim_s1} and \eqref{inner_lim_s2}, expanded in the outer region variable $t$.}

Since these regions match a composite representation of the integral can be formed by adding the inner and outer limits, subtracting the common part and then integrating the representation over the entire length. { Integrating this composite representation we find that the flow on the surface of the body is asymptotically given by
\begin{eqnarray}
\boldsymbol{u}(\boldsymbol{x_{S}}) = \boldsymbol{U}(s) &=& \int_0^1\left[  \boldsymbol{S(x_{S}-x_0}(t))\cdot\boldsymbol{\alpha}^S(t) + \boldsymbol{D(x_{S}-x_0}(t))\cdot\boldsymbol{\beta}^S(t)\right] \,dt  \notag \\
&=& \int_0^1 \left[\boldsymbol{S^{(o)}(x_{S}-x_0}(t))\cdot\boldsymbol{\alpha}^S(t) +
\boldsymbol{D^{(o)}(x_{S}-x_0}(t))\cdot\boldsymbol{\beta}^S(t) \right]\,dt \notag \\
&& +\int_0^1 \left[\boldsymbol{S^{(i)}(x_{S}-x_0}(t))\cdot\boldsymbol{\alpha}^S(s) +
\boldsymbol{D^{(i)}(x_{S}-x_0}(t))\cdot\boldsymbol{\beta}^S(s) \right]\,dt \notag\\
&& - \int_0^1 \left[\boldsymbol{S^{(o)\in(i)}(x_{S}-x_0}(t))\cdot\boldsymbol{\alpha}^S(s) +
\boldsymbol{D^{(o)\in(i)}(x_{S}-x_0}(t))\cdot\boldsymbol{\beta}^S(s) \right]\,dt \notag \\
&& + O(a) \notag\\
&=& \frac{1}{8\pi}\int_0^1\left[\frac{\boldsymbol{\alpha}^S(t)}{R_0}+ \frac{\boldsymbol{R_0R_0}\cdot\boldsymbol{\alpha}^S(t)}{R_0^3}-\frac{\mathbf{I}+\boldsymbol{e_te_t}}{|s-t|} \cdot \boldsymbol{\alpha}^S(s)\right]\,dt \notag \\
&& + \int_0^1 \left[\boldsymbol{S^{(i)}(x_{S}-x_0}(t))\cdot\boldsymbol{\alpha}^S(s) +
\boldsymbol{D^{(i)}(x_{S}-x_0}(t))\cdot\boldsymbol{\beta}^S(s) \right]\,dt \notag \\
&& + O(a), \label{SBTpre}
\end{eqnarray}
where we have explicitly substituted in the forms of the outer and common parts of the expansion. The remaining integrals over the inner region expansion can be asymptotically evaluated exactly \cite{Koens2018, Gotz2000} to find
}
\begin{eqnarray} 
8 \pi \boldsymbol{U}(s) &=& \int_0^1\left[\frac{\boldsymbol{\alpha}^S(t)}{R_0}+ \frac{\boldsymbol{R_0R_0}\cdot\boldsymbol{\alpha}^S(t)}{R_0^3}- \frac{\mathbf{I}+\boldsymbol{e_te_t}}{|s-t|} \cdot \boldsymbol{\alpha}^S(s) \right]\,dt \notag \\
&&+[ L \left(\boldsymbol{I} +\boldsymbol{e_t e_t} \right)+ \left(\boldsymbol{I} -3\boldsymbol{e_t e_t} \right)]\cdot\boldsymbol{\alpha}^S(s) {+O(a)} \label{SBT}, 
\end{eqnarray}
where $L = \ln[4s(1-s)/a^2\rho_0^2]$. The above equation is the classical slender-body theory integral equation for the unknown Stokeslet strengths $\boldsymbol{\alpha}^S(t)$. These equations have been used successfully in the modeling of several systems \cite{Lauga2016, Gaffney2011, Katsamba2020, Das2018, Higdon2006, BARTA2006, Cummins2018, Rodenborn2013a}, have been re-derived from the boundary integral representation \cite{Koens2018}, and extended to treat bodies of different shapes \cite{Borker2019, Koens2016}. Recent studies have also studied the effectiveness of these equation in different situations and bounded error on the flow from a given force to $O(a \ln a)$ \cite{Mori2021, Mori2018, Mori2020}.

Once the force per unit length along the slender body is found from Eq.~\eqref{SBT}, the resultant flow can be determined with Eq.~\eqref{flow_s}. This predicted flow has been shown to compare well with experiments of rotating helices \cite{Kim2004a} and should generally be accurate in regions close to the slender body or far away from the slender body. This is because it must asymptotically satisfy the no-slip boundary condition on the surface of the body and far away the flow can be represented by a multipole expansion with Stokeslets and source dipoles. 

Though the classic SBT equations, Eqs.~\eqref{SBT} and \eqref{flow_s}, have been very successful, the components of the integrand within Eq.~\eqref{SBT} are individually divergent. Held together these divergent factors asymptotically cancel but this makes them difficult to treat numerically. Furthermore the operator is known to encounter numerical instabilities related to eigenvalues changing sign. Hence people have created regularized slender-body theories to overcome these issues.

\subsection{Regularized slender-body theories}

The numerical issues in implementing the classical SBT equations has prompted people to create classical SBT like models using regularized Stokeslets. Two methods have been developed for this process: one where the flow is represented by a line of regularized Stokeslets and regularized source dipoles \cite{Cortez2012, Walker, HoaNguyen2014} and one where the flow is given by a line of regularized Stokeslets alone \cite{Martindale2016}. Though both are based on regularized singularities, these methods treat the non-slip boundary condition and the regularization parameter very differently.

The former method, with regularized Stokeslets and regularized source dipoles, seeks to choose the source dipole strength such that the no-slip boundary condition asymptotically applies at the surface of the body. This can be shown with a similar asymptotic expansion as done in the classical case. These methods can leave the regularization parameter free \cite{Cortez2012} or uses it to help with the expansion \cite{Walker}. In both cases, however, the regularization parameter, $\epsilon$, helps remove the numerical issue with the implementation. The flow predicted by this method therefore satisfies the boundary conditions at the surface of the body and far from it but has been shown to exhibit an additional $O(1)$ difference between it and the classical SBT that cannot be fixed with the choice of regularization parameter \cite{Ohm2021}. In practice this difference may be small and likely stems from the fact that the flow outside the body is not described by the unforced incompressible Stokes equations. 

The latter method, with a line of regularized Stokeslets alone, tries to use the regularization parameter itself to capture the thickness of the slender body \cite{Martindale2016}. As such the regularization parameter is typically chosen such that $\epsilon = k a \rho_0(s)$, where $k$ is some arbitrary constant. The strength of the Stokeslets is then set by applying the velocity condition on some effective line in space. Often this line is chosen to be the centreline of the body though it could be  applied on a line along the surface as well. This writes the problem as a line integral similar to that of classical SBT. Similarly to the previous representation, the use of regularized Stokeslets means that the flow predicted by this model will not satisfy the unforced incompressible Stokes equations outside the body for all regularizations. However the difference between the classical SBT prediction and that from this representation has not been determined.

In the remaining sections of this paper we will investigate how well a line distribution of regularized Stokeslets replicates the flow around a slender body for a generalised spherically symmetric blob and regularization parameter. This will provide us with a set of conditions that must be met in order for the far and near field flows to { asymptotically }reflect that around a slender body. Importantly we show one common power-law blob cannot satisfy the no-slip condition on the bodies surface, while compactly supported blobs do.

\section{Regularizations for the flow far from a slender body} \label{sec:force on the body} 

Far away from the slender body the flow, { predicted with SBT}, is well approximated by Eq.~\eqref{flow_s}. This is the because it is a leading order multipole representation \cite{Kim2004a}. Similarly the flow far from a line of regularized Stokeslets placed along the centreline of the body { can be approximated by expanding $\boldsymbol{S}^\epsilon(\boldsymbol{x-x_0}(t))$ in the limit $|\boldsymbol{x}| \gg 1, \epsilon$ and integrating over the length. This expansion provides
\begin{eqnarray}
\boldsymbol{u}_{\epsilon\mbox{far}}(\mathbf{x}) &=& \int_0^1 \boldsymbol{S(x-x_0}(t))\cdot\boldsymbol{\gamma}(t) \,dt + O\left(\frac{1}{|\boldsymbol{x}|^2}\right) \notag \\
 &&+  \left[ \epsilon^2\left(\int_0^\infty\frac{4\pi}{3} u^4f(u) \,du\right)\right]\int_0^1 \boldsymbol{D(x-x_0}(t)) \cdot \boldsymbol{\gamma}(t) \,dt + O\left(\frac{\epsilon^2}{|\boldsymbol{x}|^4}\right), \notag \\
\end{eqnarray}
where  $\boldsymbol{u}_{\epsilon\mbox{far}}(\mathbf{x})$ is the far field flow from the line of regularised Stokeslets and} we have used Eq.~\eqref{Stokeslet_r}. { The above approximation assumes that the functions are well behaved enough to interchange limits and integrals. We note that the source dipole term is sub-dominate to the Stokeslet in the far-field limit. However, we retain both contribution here because the Stokeslet and source dipole generate different flows, and the far-field contributions of both singularities, from the motion of a slender body, is asymptotically determined by Eq.~\eqref{flow_s}. If we compare the Stokeslet and source dipoles flows we see that} the above equation approximates Eq.~\eqref{flow_s}  if
\begin{eqnarray}
 \boldsymbol{\gamma}(t) &=& \boldsymbol{\alpha}(t), \\
2 h(s)^2\left(\int_0^\infty\frac{4\pi}{3}
u^4f(u)du\right) &=& \rho_0(s)^2, \label{far_field}
\end{eqnarray}
where we have set $\epsilon= a h(s) +O(a^2)$. The first of these conditions tells us that, in order for the flow around a line of regularized Stokeslets to replicate that of a slender body, the strength of the regularized Stokeslets must equal the force per unit length along the body. This is unsurprising. The second equation, however, provides a relationship between the regularization parameter, $\epsilon= a h(s)$, the blob used for the regularization $f(s)$, and the cross section of the body $\rho_0(s)$. It shows that, as { assumed in the existing literature}, the regularization parameter is proportional the to the radius of the cross-section along the length but suggests this proportionality constant depends on the fourth moment of the regularizing blob. Provided this moment exists, {the far field contributions from} any regularized Stokeslet can be made to {approximate the Stokeslet and source dipole flow contributions} far from a slender body.

\section{Regularization's for the flow near a slender body} \label{sec:velocity near the body}

The flow from a line of regularized Stokeslets near the boundary of the slender body can be approximated using a similar matched asymptotic expansion as that in slender-body theory. { This involves expanding the regularised Stokeslet kernel in the limit of small $a$ for an inner region, where $\xi=(t-s)/a$, $s$, and $\theta$ are fixed, and outer region, where $t$ is fixed, and then matching the solutions. The composite representation of the integral is then found by adding the outer and inner regions and subtracting the common part, just like before, to find
\begin{eqnarray}
\int_0^1 \boldsymbol{S}^\epsilon(\boldsymbol{x} - \boldsymbol{x_0}(t))\cdot\boldsymbol{\alpha}(t)dt &=&\int_0^1
\boldsymbol{S}^{\epsilon(o)}(\boldsymbol{x - x_0}(t)) \cdot\boldsymbol{\alpha}(t) dt \notag \\
&& + \int_0^1
\boldsymbol{S}^{\epsilon(i)}(\boldsymbol{x - x_0}(t))\cdot \boldsymbol{\alpha}(s) dt \notag \\
&& -\int_0^1
\boldsymbol{S}^{\epsilon(o)\in(i)}(\boldsymbol{x - x_0}(t)) \cdot \boldsymbol{\alpha}(s) dt +O(a),
\end{eqnarray}
where the superscripts $(o)$, $(i)$ and $(o)\in(i)$ again denote the expansion in the outer, inner and the outer subsequently expanded in the inner (common), respectively. In the above $\boldsymbol{x} = \boldsymbol{x_0}(s) + \boldsymbol{e_r}a\rho$ is a point close to the surface of the body and $\rho$ is a radial distance from the centreline that is of the same order as $\rho_0(s)$. Setting $\rho=\rho_0(s)$ correspond to the flow on the boundary while  $\rho>\rho_0(s)$ represents the flow close to the boundary. 

In the previous section we established that far from the line of regularised Stokeslets the kernel, $\boldsymbol{S}^\epsilon$, behaves as a line distribution of point forces and source dipoles, Eq.~\eqref{Stokeslet_r}. As a consequence in the outer region expansion of the kernel, where $t$ is fixed and $a\to0$, is asymptotically equivalent to that of a Stokeslet and a source dipole in the same limits. Hence the kernel can be expressed as
\begin{eqnarray}
\boldsymbol{S}^{\epsilon(o)}(\boldsymbol{x - x_0}(t)) \cdot \boldsymbol{\alpha}(t)  &=& \boldsymbol{S^{(o)}(x-x_0}(t))\cdot\boldsymbol{\alpha}(t) \notag \\
&& +\left[ \epsilon^2\left(\int_0^\infty\frac{4\pi}{3} u^4f(u) \,du\right)\right]  \boldsymbol{D^{(o)}(x-x_0}(t)) \cdot \boldsymbol{\alpha}(t) \notag \\
&& + O(a) \notag \\
&=& \frac{1}{8\pi}\frac{\boldsymbol{\alpha}^S(t)}{R_0}+\frac{1}{8\pi}
\frac{\boldsymbol{R_0R_0}\cdot\boldsymbol{\alpha}^S(t)}{R_0^3} +O(a),
\end{eqnarray}
which is identical to the outer expanded kernel in classical SBT. In the above we have assumed that $\epsilon$ is linearly related to $a$ as suggested by the analysis of the previous section. Since this outer kernel is the same as that found in classical SBT, the expansion of it in terms of the inner region variables will also be the be the same. The common behaviour can therefore be expressed as
\begin{eqnarray}
\boldsymbol{S}^{\epsilon(o)\in(i)}(\boldsymbol{x - x_0}(t))\cdot\boldsymbol{\alpha}^S(s) &=&  \boldsymbol{S^{(o)\in(i)}(x-x_0}(t)) \cdot\boldsymbol{\alpha}^S(s) +O(a)  \notag \\
&=& \frac{\boldsymbol{\alpha}^S(s)}{8\pi|s-t|} + 
\frac{\boldsymbol{e_te_t} \cdot\boldsymbol{\alpha}^S(s)}{8\pi|s-t|}+O(a).
\end{eqnarray}
These results can be substituted back into our composite representation for the flow near the surface of the body to find
\begin{eqnarray}
\int_0^1
\boldsymbol{S}^\epsilon(\boldsymbol{x} - \boldsymbol{x_0}(t))\cdot\boldsymbol{\alpha}(t)dt &=& \frac{1}{8\pi}\int_0^1 \left[ \frac{\boldsymbol{\alpha}(t)}{R_0} + \frac{\boldsymbol{R_0R_0}\cdot\boldsymbol{\alpha}(t)}{R_0^3}-\frac{\mathbf{I} +\boldsymbol{e_te_t} }{|s-t|} \cdot \boldsymbol{\alpha}(s) \right]dt\notag \\
&&+  \int_0^1 \boldsymbol{S}^{\epsilon(i)}(\boldsymbol{x - x_0}(t)) \cdot\boldsymbol{\alpha}(s) dt + O(a). \label{flow_r}
\end{eqnarray} }

{ The remaining inner region expansion can be deduced from Eq.~$\eqref{Stokeslet_r}$, and takes the form
\begin{eqnarray} \label{inner_reg}
 \boldsymbol{S}^{\epsilon(i)}(\boldsymbol{x -
x_0}(t)) &=&   \boldsymbol{S}^{(i)} (\boldsymbol{x -
x_0}(t)) \int_0^{\delta/h(s)} 4\pi r^2f(r)dr \notag \\
&&+ a^2h^2(s) \boldsymbol{D}^{(i)} (\boldsymbol{x -
x_0}(t)) \int_0^{\delta/h(s)}\frac{4\pi}{3} r^4f(r)dr
\notag \\
&& +\frac{2\boldsymbol{I}}{3ah(s)}\int_{\delta/h(s)}^\infty rf(r)dr +O(a),
\end{eqnarray}
where $\epsilon= a h(s)$, $\delta(t,s,a) = \sqrt{\xi^2 +
\rho^2(s)}$ and we noted that $r/\epsilon =a \delta(t,s,a)/ \epsilon  + O(a)=O(1)$.} If we consider a point on the surface $\delta=\delta_0$. The above relation expresses the inner region representation of the regularized Stokeslet in terms of the inner region representation of the Stokeslet and source dipole. { Similar inner region expansions were stated in the summary of SBT for flow on the surface of the body, Eqs.~\eqref{inner_lim_s1} and \eqref{inner_lim_s2}. The expanded form of the Stokeslet and source dipole near the surface are structurally equivalent to these but with $\delta_0,\rho_0$ replaced with $\delta,\rho$. Explicitly this gives
\begin{eqnarray}
 \boldsymbol{S}^{(i)}(\boldsymbol{x -
x_0}(t)) &=& \frac{\boldsymbol{I}}{8\pi a\delta} +
 \frac{\rho^2\boldsymbol{e_re_r}-\rho\xi(\boldsymbol{e_re_t+e_te_r})
 + \xi^2\boldsymbol{e_te_t}}{8\pi a\delta^3}, \label{st2}\\
\boldsymbol{D}^{(i)}(\boldsymbol{x - x_0}(t)) &=&
\frac{\boldsymbol{I}}{8\pi a^3\delta^3} - 3
\frac{\rho^2\boldsymbol{e_re_r}-\rho\xi(\boldsymbol{e_re_t+e_te_r})
+ \xi^2\boldsymbol{e_te_t}}{8\pi a^3\delta^5}. \label{sd2}
\end{eqnarray}}
These equations can be combined with Eqs.~\eqref{inner_reg} and \eqref{flow_r} to asymptotically determine the flow close to the surface of the slender body from a line or regularised Stokeslets.

The difference in the near body flow from a line of regularized Stokeslets and the flow around a slender body { asymptotically predicted by SBT} can be determined by subtracting Eq.~\eqref{flow_r} from  Eq.~\eqref{SBTpre}. Since the outer and common limits are the same in each case, the difference between  {the leading order contributions} at $\boldsymbol{x}=\boldsymbol{x}_0(s) + a\boldsymbol{e}_r(s)\rho(s)$ is
 {\begin{eqnarray}
 \int_0^1 \Delta \mbox{SBT}_{\epsilon} \,dt &=& 
\int_0^1 \left[\boldsymbol{S}^{(i)}(\boldsymbol{x - x_0}(t)) +
\frac{1}{2}a^2\rho_0^2(t)
\boldsymbol{D}^{(i)}(\boldsymbol{x - x_0}(t)) -\boldsymbol{S}^{\epsilon(i)}(\boldsymbol{x -
x_0}(t))\right]\,dt \cdot \boldsymbol{\alpha}(s) \notag \\
&=& \int_0^1 \boldsymbol{S}^{(i)}(\boldsymbol{x - x_0}(t))\left[1 -\int_0^{\delta/h(s)} 4\pi r^2f(r)dr \right]\,dt \cdot \boldsymbol{\alpha}(s) \notag \\
&& + \frac{1}{2}a^2\rho_0^2(s) \int_0^1  
\boldsymbol{D}^{(i)}(\boldsymbol{x - x_0}(t))\left[ 1 -2 \frac{h^2}{\rho_0^2}   \int_0^{\delta/h(s)}\frac{4\pi}{3} r^4f(r)dr\right]\,dt  \cdot \boldsymbol{\alpha}(s)\notag \\
&& - \int_0^1  \frac{2\boldsymbol{I}}{3ah(s)}\left[\int_{\delta/h(s)}^\infty rf(r)dr\right] \,dt  \cdot \boldsymbol{\alpha}(s) \notag \\
&=& \int_0^1 \boldsymbol{S}^{(i)}(\boldsymbol{x - x_0}(t))\left[\int_{\delta/h(s)}^{\infty} 4\pi r^2f(r)dr \right]\,dt \cdot \boldsymbol{\alpha}(s)  - \frac{2\boldsymbol{I}}{3ah(s)}\int_0^1  \left[\int_{\delta/h(s)}^\infty rf(r)dr\right] \,dt  \cdot \boldsymbol{\alpha}(s)\notag \\
&& + a^2 h^2(s) \int_0^1  
\boldsymbol{D}^{(i)}(\boldsymbol{x - x_0}(t))\left[  \int_{\delta/h(s)}^{\infty}\frac{4\pi}{3} r^4f(r)dr\right]\,dt  \cdot \boldsymbol{\alpha}(s)\notag \\
&& +  a^2 \left[ \frac{1}{2}\rho_0^2(s) - h^2   \int_0^{\infty}\frac{4\pi}{3} r^4f(r)dr\right] \int_0^1  
\boldsymbol{D}^{(i)}(\boldsymbol{x - x_0}(t))\,dt  \cdot \boldsymbol{\alpha}(s)
 \end{eqnarray}}
 { where $\int_0^1 \Delta \mbox{SBT}_{\epsilon} \,dt$ is the difference between the leading order contributions to the near field flows from a line of regularised Stokeslets and classical SBT. This difference needs to be O(a) for the flow around a line of regularised Stokeslets to asymptotically replicate the flow predicted by SBT to the same order. Since $\boldsymbol{\alpha}(s)$ is typically unknown, this condition must hold}
for arbitrary $\boldsymbol{\alpha}$(s). { The integrand of the above integral can be expressed as}
\begin{equation}
   -\Delta \mbox{SBT}_{\epsilon}= E_1\boldsymbol{I} + E_2\boldsymbol{e_re_r} + E_3\boldsymbol{e_te_t} +
E_4(\boldsymbol{e_re_t + e_te_r}), 
\end{equation}
where
\begin{eqnarray}
E_1&=&\frac{\delta^2}{ah^3}\int_1^\infty\left(\frac{2}{3} \varrho-\frac{1}{2} \varrho^2-\frac{1}{6}\varrho^4\right)f\left(\frac{ \varrho \delta}{h}\right)d\varrho +\frac{1}{8\pi a\delta^3} \left[ h(s)^2\left(\int_0^\infty\frac{4\pi}{3} u^4f(u)du\right) - \frac{1}2 \rho_0(s)^2 \right], \\
 E_2 &=&\frac{\rho^2}{2ah^3}\int_1^\infty(\varrho^4 -
\varrho^2)f\left(\frac{\varrho\delta}{h}\right)d\varrho-\frac{3\rho^2}{8\pi
a\delta^5} \left[ h(s)^2\left(\int_0^\infty\frac{4\pi}{3}
u^4f(u)du\right) - \frac{1}2 \rho_0(s)^2 \right],\\
E_3 &=&\frac{\xi^2}{2ah^3}\int_1^\infty(\varrho^4-\varrho^2)f\left(\frac{\varrho\delta}{h}\right)d\varrho-\frac{3\xi^2}{8\pi a\delta^5} \left[ h(s)^2\left(\int_0^\infty\frac{4\pi}{3} u^4f(u)du\right) - \frac{1}2 \rho_0(s)^2 \right],\\
E_4 &=& \frac{\rho\xi}{2ah^3}\int_1^\infty(\varrho^2 - \varrho^4)f\left(\frac{\varrho\delta}{h}\right)d\varrho+\frac{3\rho\xi}{8\pi a\delta^5} \left[ h(s)^2\left(\int_0^\infty\frac{4\pi}{3}
u^4f(u)du\right) - \frac{1}2 \rho_0(s)^2 \right].
\end{eqnarray}
{ In the above we have made the integral substitution $r = \varrho h(s)/\delta $ to change the lower bounds of the first integrals to 1 and used Eqs.~\eqref{st2} and \eqref{sd2}.}
These $E_i$ values represent the relative error between the flow { predicted by SBT} and that from the line of regularized Stokeslets { in each direction}. We require all four terms to integrate to $O(a)$ for the flow to match. { The integral of $E_4$ can be shown to be $O(a)$ by parity arguments. This is because $E_4$ is odd in $\xi$ and in the asymptotic limit $a\to 0$ the integral bounds go to $\pm \infty$.} However, the other three terms provide non-trivial conditions on $f(u)$. 

If we assume that Eq.~\eqref{far_field} is satisfied
then the conditions  $\int E_{1,2,3} \,dt = O(a)$ simplify to
\begin{eqnarray}
{ \frac{3 h^3}{2 \rho^3} \int_0^{1}\left[E_1 +\frac{1}{3}\left(E_2+E_3 \right) \right]  \,dt } &=& \int^{(1-s)/(a  \rho)}_{-s/(a \rho)}(\zeta^2+1)\left[\int_1^\infty \left( \varrho-\varrho^2\right)f\left(\frac{\varrho\rho \sqrt{1+\zeta^2} }{h}\right) \,d\varrho \right]\,d\zeta 
= O(a), \label{near_field1}\\
{ \frac{2 h^3}{\rho^3}\int_0^{1} E_2 \,dt} &  =&\int^{(1-s)/(a \rho)}_{-s/(a \rho)} \left[\int_1^\infty 
(\varrho^4-\varrho^2)f\left(\frac{\varrho\rho \sqrt{1+\zeta^2}}{h}\right)  \,d\varrho \right] \,d\zeta= O(a), \label{near_field2}\\
{\frac{2 h^3}{\rho^3} \int_0^{1} E_3 \,dt} &=&  \int^{(1-s)/(a \rho)}_{-s/(a \rho)}
\zeta^2\left[ \int_1^\infty (\varrho^4-\varrho^2)f\left(\frac{\varrho\rho \sqrt{1+\zeta^2} }{h}\right)   \,d\varrho \right] \,d\zeta =  O(a), \label{near_field3}
\end{eqnarray}
where $\zeta = \xi/\rho  =  (t-s)/(a \rho)$ and the equality needs to hold for $\rho \gtrsim \rho_0$. The above conditions provide a set of restrictions on the types of blobs, $f(s)$, that can replicate the near field flow around a slender body. { These conditions are necessary and sufficient for the line regularised Stokeslets to agree with classical SBT to O(a) near the boundary as the integrals of $E_i$ represent the difference between the two representations. These conditions therefore specify the types of regularisation's which can be used for numerical simulations to asymptotically capture the flow near a slender-body.} These conditions can be viewed as an integral transform into $\rho$ which depends on the tails of the blobs. Though the range over which these equations apply can change from blob to blob, these equations must be true when $\rho=\rho_0$. If this condition is not satisfied the non-slip boundary condition on the surface of the body cannot be satisfied. In these cases the solutions found with the line of regularized Stokeslets cannot be assumed to relate to the real solution a priori. 

\section{Testing the conditions on common blob types} \label{sec:examples}

In the previous sections we identified one condition on the regularization parameter, Eq.~\eqref{far_field}, required to { approximately} match the far field flow, and three conditions on the blob type, Eqs.~\eqref{near_field1}, \eqref{near_field2}, and \eqref{near_field3}, in order to { asymptotically} match the flow near the surface of the body. { These conditions apply in limit that the body is long and slender and were found by comparing the asymptotic behaviour around a line of regularised Stokeslets to that predicted by SBT.} Furthermore, though these latter conditions can hold for a range of $\rho$, we know they must hold when $\rho=\rho_0$ if the resultant flow is to approximate that of a slender body. In this section we apply these conditions to three common regularization blob types: power-law blobs, compact blobs and exponential blobs. In each case we identify if a line of them can be used to represent the flow from a slender body.

\subsection{Power-law blobs}

Probably the most common blob type used within the literature is the power-law type blob. These blobs are given by
\begin{equation}
    f_p({ v}) = \frac{\Gamma(n/2)}{\pi^{3/2} \Gamma((n-3)/2)} \frac{1}{({ v}^2+1)^{n/2}},
\end{equation}
where $n$ is a integer greater then 5 and $\Gamma(x)$ is the gamma function. { We remind the reader that the above blob relates to the regularising function through $f_{\epsilon} = \epsilon^{-3} f(\mathbf{r}/\epsilon)$.} The above representation reduces to the one of the most popular examples, Eq.~\eqref{pop}, when $n=7$.

\begin{figure}
\includegraphics[height=100pt]{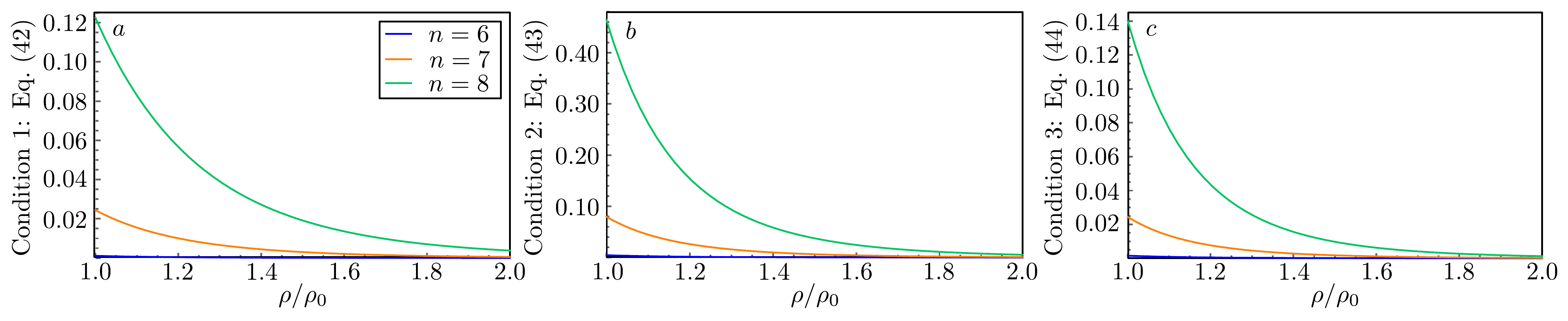}
\caption{The value of near flow conditions for power-law blobs of $n=6$ (blue), $n=7$ (orange) and $n=8$ (green). a) displays the result from condition Eq.~\eqref{near_field1} (Eq.~\eqref{pow1}), b) displays the result from condition Eq.~\eqref{near_field2} (Eq.~\eqref{pow2}), c) displays the result from condition Eq.~\eqref{near_field3} (Eq.~\eqref{pow3})} 
\label{fig:power}
\end{figure}

Inserting this blob into the far field flow condition, Eq.~\eqref{far_field}, we find
\begin{eqnarray}
h(s) =  \rho_0(s) \sqrt{\frac{n-5}{2}}. 
\end{eqnarray}
This therefore sets the regularization the blobs must use to satisfy the far field condition. { Significantly ,we see that when $n=7$, $h(s)=\rho_0(s)$. This indicates that the regularisation parameter $\epsilon$ should be taken as half the thickness of the slender-body, as is typically assumed in numerical simulations.} Similarly the near field conditions, Eqs.~\eqref{near_field1}, \eqref{near_field2} and \eqref{near_field3}, asymptotically evaluate to
\begin{eqnarray}
\int^{(1-s)/(a \rho)}_{-s/(a \rho)}(\zeta^2+1)\left[ \int_1^\infty \left( \varrho-\varrho^2\right)f_p\left(\frac{\varrho\rho \sqrt{1+\zeta^2} }{h}\right)\,d\varrho \right] \,d\zeta
 &=&  -\frac{(n-5)^n (n-2) \left((n-5)^2+4 \rho'^2\right)^{\frac{3-n}{2}}}{16 \pi  (n-3) \rho'^3}  \notag \\
 &&  -\frac{ (n-5)^n ((n-10) n-25)  F_2(\rho')}{ 2^{n+3} \rho'^{n+2}\pi  (n-1)} \notag \\
&&+\frac{(n-5)^n  F_1(\rho')}{2^n \rho'^n\pi } +\frac{ (n-5)^{n+2}  F_3(\rho')  }{2^{n+2} \rho'^{n+2}\pi  n} \notag \\
&&+O(a), \label{pow1}
\end{eqnarray}
\begin{eqnarray}
\int^{(1-s)/(a \rho)}_{-s/(a \rho)} \left[\int_1^\infty 
(\varrho^4-\varrho^2)f_p\left(\frac{r\rho \sqrt{1+\zeta^2}}{h}\right) \,d\varrho \right] \,d\zeta&=& \frac{(n-5)^{n-1} \left((n-5)^2+4 \rho'^2\right)^{\frac{5-n}{2}}}{32 \pi  \rho'^5} +O(a),   \label{pow2}\\
\int^{(1-s)/(a \rho)}_{-s/(a \rho)}
\zeta^2 \left[\int_1^\infty (\varrho^4-\varrho^2)f_p\left(\frac{\varrho\rho \sqrt{1+\zeta^2} }{h}\right)\,d\varrho \right] \,d\zeta &=& \frac{(n-5)^{n-1} \left((n-5)^2+4 \rho'^2\right)^{\frac{5-n}{2}}}{64 \pi  \rho'^5}  \notag \\
&& -\frac{\left(n-5\right)^n  F_4(\rho') }{ 2^{n+1} \rho'^n \pi  (n-3)} +O(a), \label{pow3}
\end{eqnarray}
where $\rho' = \rho/\rho_0$, $\,_2F_1(a,b;c,x)$ is the ordinary hypergeometric function, and
\begin{eqnarray}
F_1 (\rho')&=& \,_2F_1\left(\frac{n-2}{2},\frac{n-1}{2};\frac{n}{2};-\frac{(n-5)^2}{4 \rho'^2}\right), \\
F_2(\rho') &=& \,_2F_1\left(\frac{n-1}{2},\frac{n-1}{2};\frac{n+1}{2};-\frac{(n-5)^2}{4 \rho'^2}\right), \\
F_3(\rho') &=& \,_2F_1\left(\frac{n-1}{2},\frac{n}{2};\frac{n+2}{2};-\frac{(n-5)^2}{4 \rho'^2}\right), \\
F_4(\rho') &=& \,_2F_1\left(\frac{n-3}{2},\frac{n-3}{2};\frac{n-1}{2};-\frac{(n-5)^2}{4 \rho'^2}\right).
\end{eqnarray}
Clearly the above forms are not $O(a)$ to leading order and so power law blobs do not formally capture the flow. Furthermore, since the conditions are also not satisfied when $\rho'=1$, power-law blobs are incapable of capturing the no-slip boundary conditions. However, when these functions are plotted against $\rho'$ for different $n$ (Fig.~\ref{fig:power}) we find that they are small for small $n$ and increase as $n$ increases. This suggests that though these representations cannot replicate the boundary conditions directly, when $n=6$ or $n=7$ the error induced by these terms may be small. This may explain why $n=7$ blobs have seemingly been successfully used in many problems, though in practice further validation is needed to be certain that they actually represent the real flow.

Finally we note that the same conclusion can be drawn by completing the inner region expansion of the resultant regularized Stokeslet. For example when $n=7$ the regularized Stokeslet from this blob has the form of Eq.~\eqref{pop}. If we expand this in the inner limit we find

\begin{equation}
\boldsymbol{S}^{\epsilon(i)}(\boldsymbol{x - x_0}(t)) =
\frac{\boldsymbol{I}(\delta^2 + 2h^2) +
\rho^2\boldsymbol{e_re_r} - \rho\xi(\boldsymbol{e_re_t +
e_te_r}) + \xi^2\boldsymbol{e_te_t}}{8\pi
a\sqrt{\delta^2+h^2}^3} +O(a),
\end{equation}
which can then be integrated to give

\begin{equation}
 8\pi\int_0^1\boldsymbol{S}^{\epsilon(i)}(\boldsymbol{x - x_0}(t)) \,dt =
\boldsymbol{I}\left(L_2
+\frac{2\rho^2+4h^2(s)}{\rho^2+h^2(s)}\right) + \frac{2\rho^2}{\rho^2 + h^2(s)}\boldsymbol{e_re_r} +L_2\boldsymbol{e_te_t},
\end{equation}
where $L_2= \ln [4 s(1-s)/(a^2 (\rho^2+h^2))] $. The inner structure clearly has a non-zero $\boldsymbol{e_re_r}$ term that cannot be eliminated by any choice of $h$. Hence this blob cannot remove this angular dependence and cannot satisfy the boundary conditions on the surface of the slender body when $\rho=\rho_0$.  It therefore cannot be assumed a priori that the flow from a line of these regularized Stokeslets approximates the flow around a slender body. { The presence of this $\boldsymbol{e_re_r}$ term is directly related to Eq.~\eqref{pow2} containing an O(1) contribution. This is because the condition used in Eq.~\eqref{pow2} comes from the $E_2$ terms which contains the difference in the leading order terms from a line of regularised Stokeslets and SBT in the $\boldsymbol{e_re_r}$ direction. }

\subsection{Compact blobs}

Another class of blobs that can be used for regularized singularities are compact blobs. These blobs are zero beyond a certain radius and so can significantly simplify the integration. { Though such compact blobs regularise the singularity, they often generate discontinuities in the flow and so are not used as often.} For example consider the blob

\begin{equation}
    f_c({ v}) = \left\{\begin{array}{c c}
    \displaystyle \frac{15}{32 \pi} (21 { v}^2 - 11) & { v}< 1  \\
         0 & { v}\geq1
    \end{array} \right..
\end{equation}
This blob isolates the regularization to within the slender body and is chosen such that the total dipole strength is the same as in classical slender-body theory. Given these conditions we expect that the flow outside the body should solve the unforced Stokes equations and identically match the slender-body flow. This can be verified directly with the far field condition, Eq.~\eqref{far_field}, setting the regularization parameter as

\begin{equation}
    h =\rho_0,
\end{equation}
and the near field conditions becoming
\begin{eqnarray}
\int^{(1-s)/(a \rho)}_{-s/(a \rho)}(\zeta^2+1) \left[ \int_1^\infty \left( \varrho-\varrho^2\right)f_c\left(\frac{\varrho\rho \sqrt{1+\zeta^2} }{h}\right)\,d\varrho \right] \,d\zeta
&=&0, \\
\int^{(1-s)/(a \rho)}_{-s/(a \rho)} \left[ \int_1^\infty 
(\varrho^4-\varrho^2)f_c\left(\frac{\varrho\rho \sqrt{1+\zeta^2}}{h}\right) \,d\varrho \right] \,d\zeta&=&0, \\
\int^{(1-s)/(a \rho)}_{-s/(a \rho)}
\zeta^2 \left[ \int_1^\infty (\varrho^4-\varrho^2)f_c\left(\frac{\varrho\rho \sqrt{1+\zeta^2} }{h}\right)  \,d\varrho \right] \,d\zeta &=& 0,
\end{eqnarray}
because $r \rho' \sqrt{1+\zeta^2} \geq 1$ and so $f_c\left(r\rho' \sqrt{1+\zeta^2} \right) = 0$. { Again we see that the regularisation parameter $\epsilon$ equals half the body thickness, like used in simulations, however this representation also replicates all the near field conditions identically.} Hence a line of these compactly supported blobs { could be used to asymptotically determine the flow around a slender body.}

\subsection{Gaussian blobs}

The final class of blobs we will consider is Gaussian blobs.
These blobs { decay to 0 rapidly but are continuous and so can be thought of like an intermediate between the power law and the compact blobs.} Gaussian blobs have the form

\begin{equation}
    f_g({ v}) = \frac{1}{\pi^{3/2}}e^{-{ v}^2}.
\end{equation}
The far field condition for these blobs becomes

\begin{equation}
    h = \rho_0,
\end{equation}
similarly to the compact blob and { power-law blob with $n=7$} solution. However in this case the near surface conditions become
\begin{eqnarray}
\int^{(1-s)/(a \rho)}_{-s/(a \rho)}(\zeta^2+1) \left[ \int_1^\infty \left( \varrho-\varrho^2\right)f_g\left(\frac{\varrho\rho \sqrt{1+\zeta^2} }{h}\right)  \,d\varrho \right] \,d\zeta
&=& -\frac{1}{4 \pi \rho'^3} \ \int_{\rho'^2}^{\infty} \frac{e^{-u}}{u}\,du+ O(a), \notag \\ \\
\int^{(1-s)/(a \rho)}_{-s/(a \rho)} \left[ \int_1^\infty  
(\varrho^4-\varrho^2)f_g\left(\frac{\varrho\rho \sqrt{1+\zeta^2}}{h}\right)  \,d\varrho \right] \,d\zeta&=& \frac{e^{-\rho'^2}}{2 \pi \rho'^5} +O(a), \\
\int^{(1-s)/(a \rho)}_{-s/(a \rho)}
\zeta^2 \left[ \int_1^\infty (\varrho^4-\varrho^2)f_g\left(\frac{\varrho\rho \sqrt{1+\zeta^2} }{h}\right)  \,d\varrho \right] \,d\zeta &=& \frac{1}{4 \pi \rho'^5} \left(e^{-\rho'^2} - \rho'^2\int_{\rho'^2}^{\infty} \frac{e^{-u}}{u}\,du \right)+O(a). \notag \\ 
\end{eqnarray}
Again the above evaluation shows that the exponential blobs do not satisfy the conditions, however practically this error is exponentially small { in $\rho'^2$ with the largest term being $1/2\pi$ when $\rho'=1$. This again suggests that the asymptotic error from this regularisation could be small and so may still work for simulating the flow around the slender body. However it would be important to validate the simulation sufficiently beforehand. We note that the difference between the SBT model and the line of Gaussian regularized Stokeslets at the surface can be constructed from the above conditions by using them to reconstruct the $E_i$ when $\rho'=1$. }

\section{Conclusion} \label{sec:con}

In this paper we compared the { asymptotic} flow from a line of regularized Stokeslets to that from a slender body. { This was possible using SBT and performing a matched asymptotic expansion on a line of regularised stokeslets with an arbitrary blob.} We focused on the flow far from the body and near to the body surface as the SBT asymptotic representations are expected to be the most valid in these regions. We found that any line of regularized Stokeslets can { approximate} the flow far from the body provided that the regularization parameter, $\epsilon$, is equal to the thickness of the body, $a \rho_0$, multiplied by a constant that depends on the regularisation blob, $(8 \pi  \int_0^{\infty} u^4 f(u) \,du/3)^{-1/2} $, Eq.~\eqref{far_field}. { This constant can be used for the unknown proportionality constant used in numerical simulations.} The {asymptotic} flow near the surface of a slender body, however, could only be captured by specific regularisation blobs.

The regularized Stokeslets which are capable of {asymptotically} describing the flow near the surface of a slender body are found to satisfy three conditions, Eqs.~\eqref{near_field1}, \eqref{near_field2}, and \eqref{near_field3}. These conditions depend on the tails of the regularization blob and are required to apply over a region. Importantly these conditions must be met at the surface of the slender body if the drag from the line of regularized Stokeslets is to correctly predict the same force and torque on the body from a given motion. If it does not the representation cannot be assumed to describe the flow around a slender body. { This restricts the types of regularization blobs that could be used when simulating a slender-body with a line of regularised Stokeslets.}

Finally we tested these conditions on three blob types: power-law blobs, compact blobs, and Gaussian blobs. While all could satisfy the far field condition without issue we found that power-law blobs and Gaussian blobs could not satisfy the near field conditions anywhere. As such these blobs cannot match the correct non-slip condition on the bodies surface and so models using these regularization's may not describe the desired system. Plots of the error suggested that these differences are small and so may not make a large difference practically. The compact blobs investigated were found to satisfy all the conditions without issue. { This means that, out of the three blobs considered, the compact blob is the best suited for modelling the behaviour of slender-bodies with a line of regularised Stokeslets. }

In this article we restricted ourselves to spherically symmetric regularizations and focused on slender-bodies with circular cross-sections. It would be of interest to extend the results to non-spherically symmetric regularizations and slender-bodies with non-circular cross-sections to see how the conditions change. Furthermore slender-body theory has also been used in potential flow and diffusion problems and so the method could be extended to investigate the accuracy of a line of singularities for a slender body in those systems. Finally we note that the only regularization we found that satisfied all the conditions was a compact blob and so it would be of interest to look for any non-compact examples which work.

\acknowledgments{ LK was funded by Australian Research Council (ARC) under the Discovery Early Career Research Award scheme (grant agreement DE200100168). The authors also thank Eric Lauga for useful discussions and advice.}

\bibliographystyle{ieeetr}
\bibliography{references}

\end{document}